\documentclass[a4paper]{article}

\usepackage{INTERSPEECH2022}

\usepackage{algorithm}
\usepackage[noend]{algpseudocode}
\usepackage{arydshln}
\usepackage[multiple]{footmisc}
\usepackage{tablefootnote}
\usepackage{xurl}
\usepackage{booktabs}
\usepackage[pagebackref,hidelinks]{hyperref}

 \hypersetup{
     colorlinks=true,
     linkcolor=blue,
     filecolor=blue,
     citecolor = black,      
     urlcolor=cyan,
     }

\renewcommand*{\backref}[1]{}
\renewcommand*{\backrefalt}[4]{[{%
    \ifcase #1 Not cited.%
          \or Cited on page~#2.%
          \else Cited on pages #2.%
    \fi%
    }]}

\title{SHAS: Approaching optimal Segmentation for End-to-End Speech Translation}
\name{Ioannis Tsiamas, Gerard I. Gállego, José A. R. Fonollosa, Marta R. Costa-jussà}
\address{TALP Research Center, Universitat Politecnica de Catalunya, Barcelona}
\email{\{ioannis.tsiamas,gerard.ion.gallego,jose.fonollosa,marta.ruiz\}@upc.edu}

\begin{document}

\maketitle

\begin{abstract}
    Speech translation models are unable to directly process long audios, like TED talks, which have to be split into shorter segments. Speech translation datasets provide manual segmentations of the audios, which are not available in real-world scenarios, and existing segmentation methods usually significantly reduce translation quality at inference time. To bridge the gap between the manual segmentation of training and the automatic one at inference, we propose Supervised Hybrid Audio Segmentation (SHAS), a method that can effectively learn the optimal segmentation from any manually segmented speech corpus. First, we train a classifier to identify the included frames in a segmentation, using speech representations from a pre-trained wav2vec 2.0. The optimal splitting points are then found by a probabilistic Divide-and-Conquer algorithm that progressively splits at the frame of lowest probability until all segments are below a pre-specified length. Experiments on MuST-C and mTEDx show that the translation of the segments produced by our method approaches the quality of the manual segmentation on 5 language pairs. Namely, SHAS retains 95-98$\%$ of the manual segmentation's BLEU score, compared to the 87-93$\%$ of the best existing methods. Our method is additionally generalizable to different domains and achieves high zero-shot performance in unseen languages.
\end{abstract}

\noindent\textbf{Index Terms}: speech translation, audio segmentation

\setlength{\parindent}{12pt}

\section{Introduction}

    The traditional approach to speech translation (ST) has been to combine an automatic speech recognition (ASR) module, with a machine translation (MT) module, which is known in the literature as \textit{cascaded}. Recent advances in deep learning (e.g. Transformer \cite{attention}), coupled with an increased availability of ST corpora (e.g. MuST-C \cite{mustc}, mTEDx \cite{mtedx}), have enabled ST systems to be trained \textit{end-to-end}, thus bypassing shortcomings of the cascaded approach (e.g. error propagation, slow inference) and achieving very competitive results \cite{cascade_vs_end2end}. \par
    
    For MT, a document can be segmented by splitting on strong punctuation and for ASR, segmentation is of lower importance due to the more local context that is required for the task, and thus overlapping windows are usually sufficient. Segmentation for cascaded ST can be achieved by generating the ASR transcript, monolingual translation to restore punctuation, and then splitting on punctuation before applying MT \cite{matusov2007,apptek2018}. On the contrary, for end-to-end ST, proper segmentation is both important due to the significance of context, and non-trivial due to the absence of linguistic features. ST corpora provide a manual segmentation (for both training and testing sets), derived from punctuation in the transcripts, and thus the task of audio segmentation is usually neglected in most research settings. But in real-world scenarios, the manual segmentation is not available, and an automatic segmentation has to be derived from the audio. Recent evaluation campaigns of IWSLT \cite{iwslt2020, iwslt2021} simulate this setting and have shown light on the importance of segmentation in ST, where top-performing submissions used their own segmentation algorithms and achieved significantly better results. \par
    
    Segmentation methods for end-to-end ST are either \textit{length-based}, where the audio is split at fixed points, \textit{pause-based}, where the audio is split at pauses detected usually by \textit{Voice Activity Detection} (VAD) tools \cite{vad}, or \textit{hybrid} \cite{srpol2020}, where the segments are created by taking into account both the length of the resulting segments and the audio content. \cite{beyond_vad} compared different segmentation methods and proposed a hybrid method that provides more control over the resulting segment's length. They show that the proposed method provides improvements over the traditional length- and pause-based approaches, but the gap between automatic and manual segmentation still remains considerably large. Namely at 3-4 BLEU points behind the manual segmentation and retaining only 88.5$\%$ of its score \cite{beyond_vad}. \par
    
    In this research, we aim to bridge this gap, by proposing \textit{Supervised Hybrid Audio Segmentation} (SHAS), a new method that learns the manual segmentation from a labeled speech corpus. First, our method trains a classifier that predicts whether a frame should be included in the segmentation, utilizing contextual speech representations extracted from a pre-trained wav2vec 2.0 \cite{wav2vec2.0}. At inference, a probabilistic version of the \textit{Divide-and-Conquer} (pDAC) algorithm \cite{srpol2020}, uses the predictions of the classifier, and progressively splits the audio at the frames of lowest probability until all segments are below a pre-specified length. We carry out experiments on 5 language directions of MuST-C and mTEDx, and show that the SHAS retains on average 95.7$\%$ of the original BLEU. A version of our method with a multilingual classifier can achieve further improvements, retaining on average 96.3$\%$ of the original BLEU, while even decreasing the gap with manual segmentation down to 0.5 BLEU for a specific language pair. Furthermore, we find that SHAS can be transferred to different domains and has very high zero-shot performance in languages not seen during the training of the classifier. Our code and models are publicly available\footnote{\scriptsize \href{https://github.com/mt-upc/SHAS}{https://github.com/mt-upc/SHAS}}.

\section{Background}

\setlength{\parindent}{0pt}

    \textbf{wav2vec 2.0 models.} This is a family of speech encoders, pre-trained by self-supervision on unlabeled speech, and are composed of two main blocks. A convolutional feature extractor that processes the raw audio waveform and a Transformer encoder that extracts contextualized representations. After self-supervision, it can be fine-tuned to downstream tasks like ASR. Its multilingual version, XLS-R \cite{xls-r} has been pre-trained on 128 languages using 436k hours of speech data. \par
    
    \textbf{Audio Segmentation methods.} Length-based methods split the audio into fixed-length segments and their advantage is that no external system is required for segmentation, but segments will probably not resemble proper sentences, since the acoustic properties of the audio are disregarded. On the contrary, pause-based methods split the audio using only its acoustic features, usually by applying VAD tools. They classify a frame of audio as speech or non-speech and split when enough consecutive frames are classified as non-speech, essentially performing silence removal. The absence of syntactic information usually leads to performance drops \cite{sinclair2014}. Finally, hybrid methods take into account both the resulting segment's length and the acoustic features of the audio. Two different algorithms can be used for hybrid segmentation, the \textit{Divide-and-Conquer} (DAC) \cite{srpol2020} and the \textit{Streaming} (STRM) \cite{beyond_vad}. DAC progressively splits the audio on the longest detected pauses, until all segments are below a \textit{max-segment-length (max)} parameter. STRM additionally utilizes a \textit{min-segment-length (min)} parameter to split on the longest detected pause between the \textit{min} and \textit{max} seconds in an audio stream of \textit{max} seconds. If a pause does not exist in that range, the whole stream is used as a segment. STRM will produce longer segments and has the advantage that the full audio is not required in advance. \cite{upc2021} used a pre-trained wav2vec 2.0 instead of VAD, as a pause detector, where a frame is classified as non-speech if wav2vec 2.0 predicts a padding token.
    
    \setlength{\parindent}{12pt}

\section{Methodology}

\begin{figure}[h!]
        \centering
        \includegraphics[width=0.8\columnwidth]{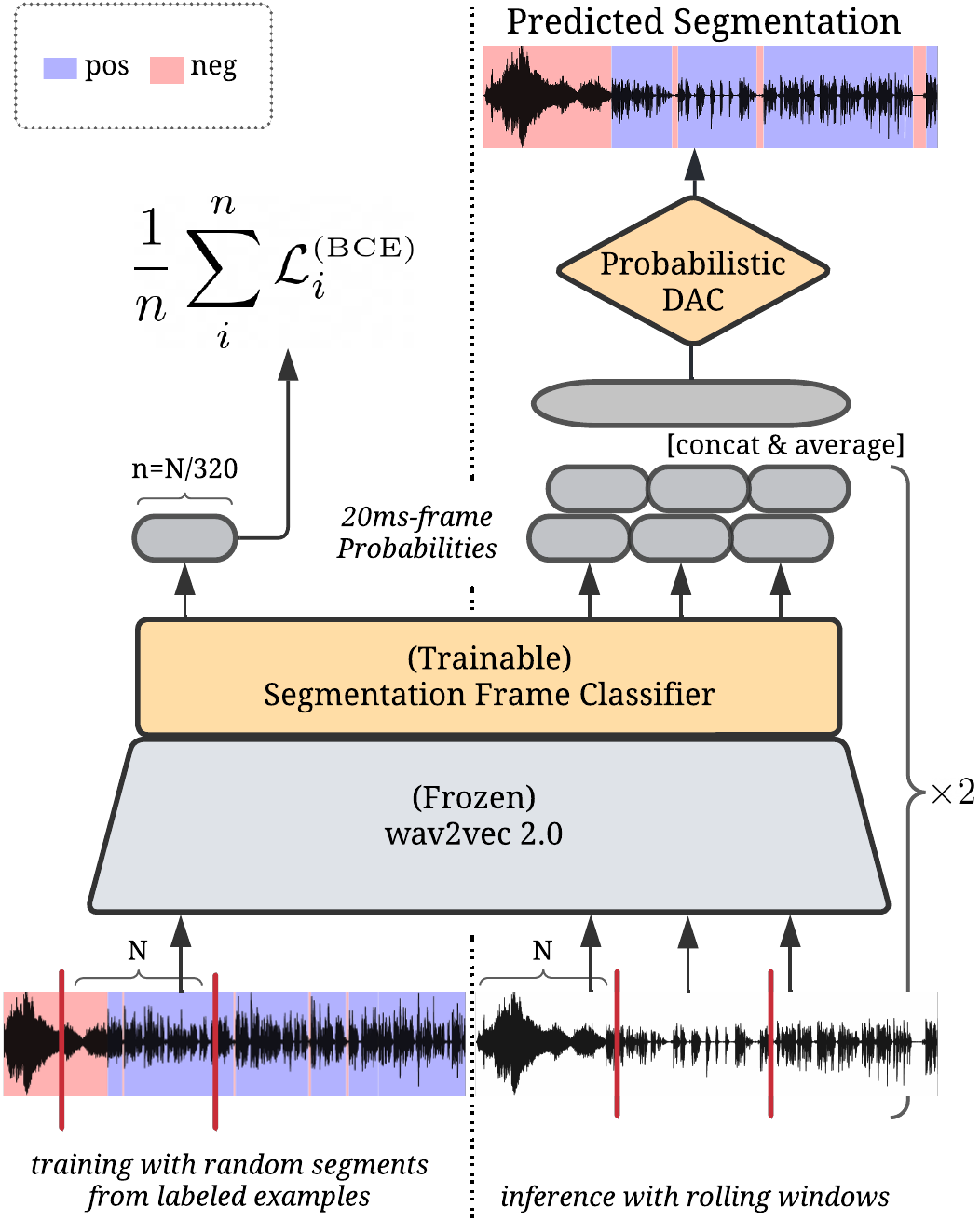}
        \caption{Supervised Hybrid Audio Segmentation (SHAS). \newline Left: Training procedure. Right: Segmentation at inference.}
        \label{fig:method}
    \end{figure}

    Our proposed method, SHAS (Figure \ref{fig:method}), generates a segmentation by applying a probabilistic Divide-and-Conquer (pDAC) algorithm on the predictions of a learned Segmentation Frame Classifier (SFC). \par
    
    \setlength{\parindent}{0pt}

    \textbf{Segmentation Frame Classifier.} This module is trained in a supervised manner and can effectively learn the splitting frames of the manual segmentation. To train the classifier we use data from ASR or ST corpora that have been manually segmented. Specifically, given a speech corpus, we ignore the transcriptions/translations and use only their segment boundaries to label each frame of audio as positive or negative, on whether it is included in a segment. For the cases where the end-frame of a segment coincides with the start-frame of the next, we explicitly label it as negative, to indicate a split. We create training examples of length $N$, by randomly segmenting each waveform. The random segmentation differs at each training epoch. The waveform $x \in \mathbb{R}^N$ of a random segment is normalized and passed through a pre-trained wav2vec 2.0 speech encoder that extracts contextualized representations $H \in \mathbb{R}^{n \times d}$, where $n=N/320$, due to the convolutional feature extractor of wav2vec 2.0. We keep the parameters of wav2vec 2.0 fixed during training, and thus we found it beneficial to further process the representation with a Transformer encoder layer, before passing them through a linear sigmoid layer that maps them to a sequence of binary probabilities $p_i \in [0,1]$, where $i=1,\dots,n$. The parameters of the Transformer and the linear layer are trained using the binary cross-entropy loss, weighted towards the negative examples, to account for class imbalance. At inference time, given an unlabeled audio waveform, we obtain the probabilities for each frame with a non-overlapping fixed rolling window of length $N$. To minimize context loss, we perform rolling inference with 2 different offsets and then average the resulting probabilities of each frame. \par
    
    \begin{algorithm}[h!]
        \footnotesize
        \caption{Probabilistic DAC}\label{alg:prob}
        \begin{algorithmic}[1]
            \Procedure{recursive\_split}{sgm}
                \If{$\textbf{len}(sgm) < max$}
                    \State \textbf{append} $sgm$ \textbf{to} $segments$ 
                \Else
                    \State $j \leftarrow 0$
                    \State $indices \leftarrow$ \textbf{argsort} $probs[sgm]$
                    \While{True}
                        \State $sgm\_a,sgm\_b \leftarrow$ \textbf{split} $sgm$ \textbf{at } $indices[j]$
                        \State $sgm\_a \leftarrow$ \textbf{trim}$(probs[sgm\_a], thr)$
                        \State $sgm\_b \leftarrow$ \textbf{trim}$(probs[sgm\_b], thr)$
                        \If{$\textbf{len}(sgm\_a) > min$ \textbf{and} $\textbf{len}(sgm\_b) > min$}
                            \State \Call{recursive\_split}{sgm\_a}
                            \State \Call{recursive\_split}{sgm\_b}
                            \State \textbf{break}
                        \EndIf
                        \State $j \leftarrow j + 1$
                    \EndWhile
                \EndIf
            \EndProcedure

            \Procedure{probabilistic\_dac}{$probs,max,min,thr$}
                \State $segments \leftarrow$ \text{empty List}
                \State $sgm \leftarrow$ \text{Tuple}[$0$, \textbf{len}($probs$)] \Comment{init single segment}
                \State \Call{recursive\_split}{$sgm$}
                \State \textbf{return} $segments$
            \EndProcedure
        \end{algorithmic}
    \end{algorithm}
    
    \textbf{Probabilistic DAC.} After obtaining the probabilities for each frame of the audio, we use the pDAC segmentation algorithm (Alg.\ref{alg:prob}) to obtain a segmentation. Instead of splitting at long pauses, as in \cite{srpol2020}, pDAC splits at the points of lowest probability. The algorithm is parameterized by a \textit{max-segment-length (max)} to control the resulting segments length, \textit{min-segment-length (min)} to avoid very small noisy segments, and a \textit{threshold (thr)} to trim the ends of a segment that are classified as excluded. pDAC progressively splits at the frame of lowest probability, until all segments are shorter than \textit{max}. After a split, the resulting segments are trimmed to the first and last frames $i, j$ with $p_i, p_j > thr$. The split will not take place unless the length of the resulting segments is above \textit{min}, and the frame of the next lowest probability will be selected. Although not included for simplicity in Alg.\ref{alg:prob}, the split will happen at the frame of lowest probability if none of the frames satisfy the \textit{min} conditions.

    \setlength{\parindent}{12pt}

\section{Experimental Setup} \label{sec:experimental_setup}

\setlength{\parindent}{0pt}

    \textbf{Data}. We conducted experiments on five language pairs: from English to German (en-de) and to English from Spanish (es-en), French (fr-en), Italian (it-en), Portuguese (pt-en). We used MuST-C en-de for training an English SFC, while Europarl-ST en-de \cite{europarl} was used only for testing purposes. For the non-English source languages, we trained the classifiers with the ASR data of mTEDx (es-es, fr-fr, it-it, pt-pt), and used the ST valid and test splits for development and testing purposes. MuST-C and mTEDx are based on TED talks, while Europarl-ST is based on talks from the European Parliament. All audio data were converted to mono and sampled at 16kHz. \par
    
    \textbf{Segmentation Frame Classifiers}. We trained five monolingual SFC models on English (en), Spanish (es), French (fr), Italian (it), and Portuguese (pt), and a multilingual one, on all source languages. All models were trained with random segments of $N=320k$ audio samples (20 secs). We used the 300m parameter version of the XLS-R model\footnote{\scriptsize \href{https://huggingface.co/facebook/wav2vec2-xls-r-300m}{https://huggingface.co/facebook/wav2vec2-xls-r-300m}} from the Transformers library \cite{huggingface}, which has 24 layers and dimensionality 1024. We used the representations of the 14th layer of XLS-R as inputs to the SFC, which was found in preliminary experiments to increase the translation quality in the development set of MuST-C en-de by 1.5 BLEU. The Transformer encoder of the classifier has a single layer, 1024 model dimension, feed-forward dimension of 2048, 8 heads, pre-layer normalization \cite{prelayer_norm}, GELU activation \cite{gelu}, and 0.1 dropout. Another layer normalization and 0.1 dropout were used before mapping to probabilities with a linear sigmoid layer. All models were trained for 8 epochs with Adam \cite{adam} and an initial learning rate of $2.5 \cdot 10^{-4}$, which decays with cosine annealing. The batch size was set at 14 examples and the update frequency at 20. After training, we chose the best checkpoint according to the quality of the translations produced by the segmentation on the development sets. For evaluation, we perform 2 rolling window inferences with different fixed-length segmentations of an audio, both with the training length $N$ of 320k samples, and average them to obtain the segmentation frame probabilities of the whole audio. \par
    
    \textbf{Segmentation methods and algorithms.} The development sets were used to find the optimal parameters of each segmentation method. For the pDAC of our the SHAS method, we tuned the \textit{max} parameter in the range of (4, 40) seconds. For all language directions, the parameter \textit{min} was set to 0.2 seconds, and the \textit{thr} parameter to 0.5. For pause-based segmentation, we used WebRTC's VAD\footnote{\scriptsize \href{https://github.com/wiseman/py-webrtcvad}{https://github.com/wiseman/py-webrtcvad}}, which splits the audio when 90$\%$ of consecutive frames do not include speech, and tuned the \textit{frame-length} parameter in (10, 20, 30) ms and the \textit{aggressiveness} parameter in (1, 2, 3), where higher values mean more aggressive splits. For the length-based and hybrid-DAC methods, we tuned the \textit{max} in range of (4, 40), while for the hybrid-STRM ones, we tuned the \textit{min} in the range of (0, \textit{max}), where \textit{max} is the optimal parameter for the corresponding hybrid-DAC method. For the hybrid-W2V, we do inference with fine-tuned wav2vec 2.0 models from the Transformers library\footnote{\scriptsize \href{https://huggingface.co/facebook/wav2vec2-large-960h-lv60-self}{wav2vec2-large-960h-lv60-self}, \href{https://huggingface.co/jonatasgrosman/wav2vec2-large-xlsr-53-spanish}{wav2vec2-large-xlsr-53-es}, \newline \href{https://huggingface.co/jonatasgrosman/wav2vec2-large-xlsr-53-french}{wav2vec2-large-xlsr-53-fr}, \href{https://huggingface.co/jonatasgrosman/wav2vec2-large-xlsr-53-italian}{wav2vec2-large-xlsr-53-it}, \href{https://huggingface.co/jonatasgrosman/wav2vec2-large-xlsr-53-portuguese}{wav2vec2-large-xlsr-53-pt}}, and then apply the DAC, as in \cite{upc2021}, or, similarly, the STRM algorithm.
    
    \textbf{Speech Translation Models}. For ST we used the \textit{Joint Speech-to-Text} \cite{joint-s2t} models from fairseq \cite{fairseq}. These are Transformer encoder-decoders that can take both speech (ST) and text (MT) as inputs, and have the top layers of the encoder shared between the two modalities. Apart from the negative log-likelihood, online knowledge distillation from the MT task is used to guide the ST task, and cross-attentive regularization is applied to the representations of the encoder to bridge the gap between the two modalities. For en-de translations we used the model\footnote{\scriptsize \href{https://github.com/pytorch/fairseq/blob/main/examples/speech_text_joint_to_text/docs/ende-mustc.md}{fairseq/blob/main/examples/speech\_text\_joint\_to\_text/docs/ende-mustc.md}} trained on MuST-C en-de, that has 12 encoder and 6 decoder layers, dimensionality of 512, and feed-forward dimension of 2048. The inputs to the model during inference are 80d log mel-filterbank features computed every 10ms with a 25ms window, while global channel mean and variance normalization is also applied. For the x-en translations we used the multilingual model\footnote{\scriptsize \href{https://github.com/pytorch/fairseq/blob/main/examples/speech_text_joint_to_text/docs/iwslt2021.md}{fairseq/blob/main/examples/speech\_text\_joint\_to\_text/docs/iwslt21.md}}, which is trained on the es-en, es-es, fr-en, fr-es, fr-fr, it-it, pt-en and pt-pt pairs of mTEDx, has 24 encoder and 12 decoder layers, dimensionality of 768 and feed-forward dimension of 3076. The inputs to the model during inference are audio waves sampled at 16kHz. The target vocabularies are learned by SentencePiece \cite{sentencepiece}, with a size of 10,000 for the en-de model and 64,000 (shared) for the multilingual one. For both these models, only speech is used during inference, and decoding is done with a beam search of size 5. \par
    
    \textbf{Evaluation}. After creating a new segmentation of a test set, with either SHAS or other segmentation methods, we translate the segments using an ST model, align the translations with the references using mwerSegmenter \cite{mwersegmenter}, and then compute the BLEU scores \cite{bleu} with SacreBLEU \cite{sacrebleu}.

\setlength{\parindent}{12pt}

\section{Results}

    \begin{table*}[th]
        \caption{BLEU scores of SHAS, manual segmentation, and other methods. In parenthesis is the percentage of manual BLEU score retained. \textbf{(i)}: Main results on MuST-C en-de tst-COMMON and mTEDx x-en test. \textbf{(ii)}: Cross-domain results on Europarl-ST test.}
        \label{tab:main_results}
        \centering
        \resizebox{\textwidth}{!}{
        \begin{tabular}{lcccccccc}
            \cmidrule[1pt]{1-7} \cmidrule[1pt]{9-9}
            \textbf{Segm. Methods} & \textbf{en-de}        & \textbf{es-en}        & \textbf{fr-en}        & \textbf{it-en}        & \textbf{pt-en}        & \textbf{Average}      & \text{} & \textbf{Europarl en-de} \\ \cmidrule{1-7} \cmidrule{9-9}
            Manual & 26.99 (100.)          & 31.94 (100.)          & 36.69 (100.)          & 27.15 (100.)          & 34.88 (100.)          & 31.53 (100.)          &\text{} & 28.83 (100.)            \\ \cmidrule{1-7} \cmidrule{9-9}
            Length-based (fixed)           & 22.34 (82.8)          & 27.71 (86.8)          & 30.57 (83.3)          & 23.66 (87.1)          & 30.21 (86.6)          & 26.90 (85.3)           &\text{} & 22.35 (80.3)            \\
            Pause-based (VAD)          & 22.78 (84.4)          & 27.03 (84.6)          & 30.01 (81.8)          & 21.77 (80.2)          & 26.58 (76.2)          & 25.63 (81.3)          &\text{} & 18.58 (66.8)            \\
            Hybrid VAD-DAC         & 24.19 (89.6)          & 29.38 (92.0)          & 31.85 (86.8)          & 24.35 (89.7)          & 30.92 (88.6)          & 28.14 (89.2)          &\text{} & 25.06 (90.1)            \\
            Hybrid VAD-STRM        & 23.75 (88.0)          & 29.54 (92.5)          & 31.79 (86.6)          & 24.72 (91.0)          & 31.15 (89.3)          & 28.19 (89.4)          &\text{} & 24.08 (86.5)            \\
            Hybrid W2V-DAC         & 24.49 (90.7)          & 29.69 (93.0)          & 33.48 (91.3)          & 24.82 (91.4)          & 32.48 (93.1)          & 28.99 (91.9)          &\text{} & 24.92 (89.5)            \\
            Hybrid W2V-STRM        & 24.47 (90.7)          & 29.50 (92.4)           & 33.40 (91.0)           & 25.24 (93.0)          & 32.35 (92.7)          & 28.99 (91.9)          &\text{} & 23.27 (83.6)            \\ \cmidrule{1-7} \cmidrule{9-9}
            SHAS               & \textbf{25.67 (95.1)} & 30.50 (95.5)           & 35.08 (95.6)          & 26.38 (97.2)          & 33.20 (95.2)           & 30.17 (95.7)          &\text{} & \textbf{26.40 (94.9)}   \\
            $\hookrightarrow$ \text{Multilingual}         & 25.61 (94.9)          & \textbf{30.82 (96.5)} & \textbf{35.28 (96.2)} & \textbf{26.56 (97.8)} & \textbf{33.53 (96.1)} & \textbf{30.36 (96.3)} &\text{} & 26.24 (94.3)            \\ \cmidrule[1pt]{1-7} \cmidrule[1pt]{9-9}
        \end{tabular}
        }
        \end{table*}

    In the left part of Table \ref{tab:main_results}, we compare the BLEU scores obtained from translating the segmentation of SHAS, to those of other existing methods. The BLEU scores of the manual segmentation's translations\footnote{\scriptsize minor differences due to fairseq version used for generation} serve as an upper bound to the quality of the segmentation. We compute the average BLEU of each method across the five tested language pairs. \par
    
    We observe that the SHAS is producing high-quality segmentations, allowing the ST models to retain more than 95$\%$ of the manual BLEU in all language directions, and 95.7$\%$ on average. With respect to the other segmentation methods, we obtain improvements of 4.5 BLEU from the classical pause-based approach and 2 BLEU from the hybrid approaches of \cite{srpol2020} and \cite{beyond_vad}. Compared to the hybrid method of \cite{upc2021}, which also employs pre-trained wav2vec 2.0 models, SHAS achieves an increase of 1.2 in BLEU or 3.8$\%$ closer to the BLEU of the manual segmentation. Additional improvements can be observed for all the language pairs, apart from the most high-resourced en-de, by using a multilingual SFC. Especially, for it-en SHAS reduces the gap with the manual segmentation to only 0.5 BLEU. \par

    We find that the pause-based method is under-performing, retaining on average 81.3$\%$ of the manual BLEU, and that even the simpler length-based approach can achieve better results (85.3$\%$). This is likely caused by the less accurate VAD predictions on non-English audio, since the picture is different for the en-de pair, where pause-based works better. For the hybrid methods, we confirm the relative performance of the Hybrid VAD-STRM method, which, for MuST-C en-de, retains 88.5$\%$ BLEU in \cite{beyond_vad} and 88$\%$ in Table \ref{tab:main_results}. We find on average no difference between the DAC and STRM methods, although the latter one should be preferable for its applicability since it does not require the full audio to perform the segmentation. Finally, wav2vec 2.0 models (W2V) are better pause predictors compared to VAD, aiding the hybrid algorithms to produce better segments, with improvements of 0.8 BLEU on average. \par

    \begin{table}[th]
        \caption{Zero-shot BLEU of SHAS, with classifiers (SFC) trained on the language of the index. With bold are the best and underlined are the supervised ones. Avg-zs is the average BLEU of the non-supervised scores.}
        \vspace{-0.2cm}
        \label{tab:zero_shot}
        \centering
        \resizebox{0.9\columnwidth}{!}{
        \begin{tabular}{lcccccc}
            \toprule
            \textbf{SFC} & \textbf{en-de} & \textbf{es-en} & \textbf{fr-en} & \textbf{it-en} & \textbf{pt-en} & \textbf{Avg-zs} \\ \midrule
            en                     & \textbf{\underline{25.67}} & 22.26          & 33.62          & 14.34          & 26.12          & 24.08           \\
            es                     & 21.73          & \underline{30.50}          & 32.78          & 25.78          & 33.01          & 28.32           \\
            fr                     & 23.71          & 30.28          & \textbf{\underline{35.08}} & \textbf{26.41} & 32.64          & 28.26           \\
            it                     & 22.22          & 30.37          & 34.43          & \underline{26.38}          & 32.82          & \textbf{29.96}  \\
            pt                     & 23.02          & \textbf{30.84} & 34.11          & 26.16          & \textbf{\underline{33.20}} & 28.53           \\ \bottomrule
        \end{tabular}
        }
    \end{table}
    
    \begin{figure}[th]
        \centering
        \includegraphics[width=\columnwidth]{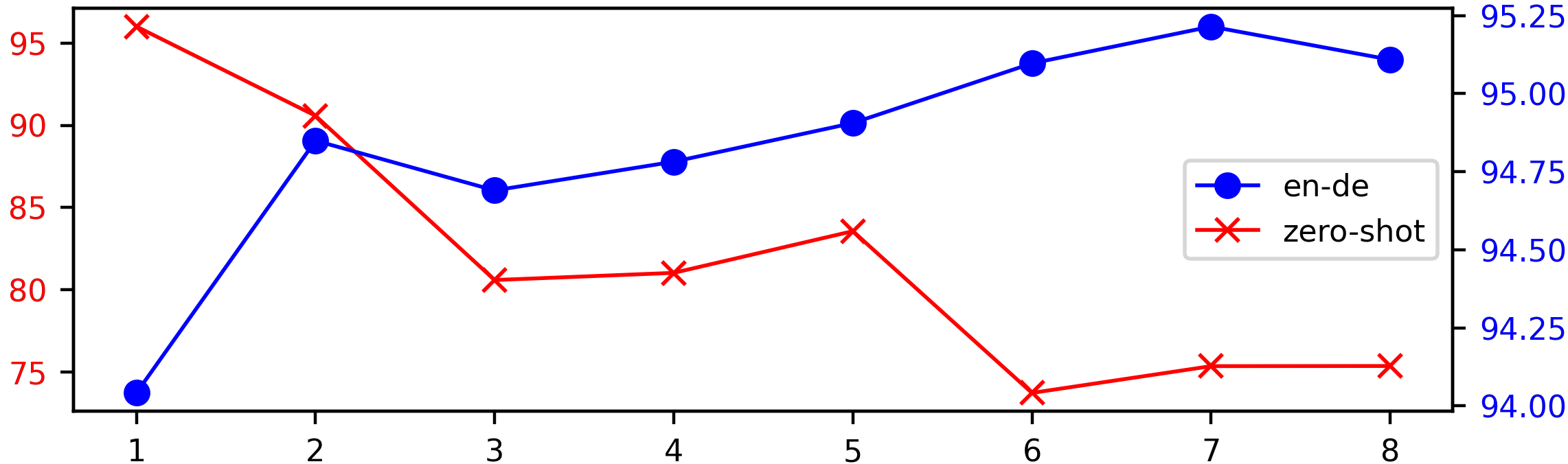}
        \caption{SHAS with English SFC for the supervised and zero-shot directions. x-axis: training epochs. y-axis: percentage of manual BLEU.}
        \label{fig:en-zero-shot}
    \end{figure}

    To investigate the generalizability of SHAS, we performed experiments on cross-domain applicability to Europarl-ST and zero-shot performance to other languages. We applied SHAS, with a classifier trained on MuST-C en-de, to the test set of Europarl-ST en-de. No hyperparameter is fine-tuned on this specific domain, and all segmentation methods are applied with exactly the same configurations that were used for MuST-C en-de. In the right part of Table \ref{tab:main_results}, we observe that the SHAS can be applied successfully to an unseen domain, where it achieves 26.4 BLEU or 94.9$\%$ of the manual segmentation's BLEU, compared to the 25.06 (90.1$\%$) of the best existing approach. Following, in Table \ref{tab:zero_shot} we apply the proposed method with monolingual SFC models trained on the five source languages, to the test sets of the other language pairs. The zero-shot BLEU scores are very high and comparable to the ones of the supervised directions in most cases. This is a direct consequence of the XLS-R backbone of the models, which has been pre-trained on all five languages. Training a classifier, with only a single Transformer encoder, on the representations of XLS-R does not result in loss of generalization, and all the models, apart from the English one, achieve on average very high zero-shot performance. These results indicate that we can obtain high-quality segmentations on many languages, by having manually segmented speech data on only one. Further research will investigate this effect on low-resource languages included in the XLS-R pre-training. The lower zero-shot performance of SHAS with an English classifier is likely explained by the larger corpus that was used (450h vs 100-200h), which eventually causes the model to become really specialized. In Figure \ref{fig:en-zero-shot}, we observe the trade-off between specialization in the supervised direction and loss of generalization in the zero-shot ones. In the early stages of training, SHAS is still generalizable but is not optimal yet for the supervised one. \par
    
    In Table \ref{tab:ablations} we investigate the importance of different components of SHAS on MuST-C en-de. Our method can also work on streams of audio, where a probabilistic version of the STRM algorithm \cite{beyond_vad} is 0.1 BLEU worse than the pDAC. A more compact version of SHAS, where we used the output of the feature extractor of XLS-R (and fine-tuned its parameters) instead of using the output of the (frozen) 14th layer, can also achieve competitive results, lagging 0.4 BLEU behind. Furthermore, we observe that not using a Transformer layer in the SFC has a negative impact of 1.2 BLEU and that increasing the number of Transformer layers to 2 does not bring further improvements. In general, these results show that the configuration of SHAS is optimal for an offline setting, but faster and more compact versions still produce better results than existing segmentation methods, and could thus be applied in online settings, where latency and model size are more important.
        
    \begin{table}[th]
        \caption{Ablations on MuST-C en-de. Params are in millions.}
        \label{tab:ablations}
        \centering
        \resizebox{\columnwidth}{!}{
        \begin{tabular}{lcccc}
            \toprule
            \textbf{Segm. Methods}        & \textbf{\begin{tabular}[c]{@{}c@{}}Full\\ Audio\end{tabular}} & \textbf{\begin{tabular}[c]{@{}c@{}}Train\\ params\end{tabular}} & \textbf{\begin{tabular}[c]{@{}c@{}}Infer\\ params\end{tabular}} & \textbf{BLEU} \\
            \midrule
            SHAS                    & $\checkmark$ & 8.4              & 208.5            & 25.67 (95.1) \\
            \midrule
            $\hookrightarrow$ pSTRM    & $\times$ & 8.4              & 208.5            & 25.57 (94.7) \\
            $\hookrightarrow$ FT XLS-R Feat. Extr.   & $\checkmark$ & 25.7             & 25.7             & 25.26 (93.6) \\
            $\hookrightarrow$ 0 Transf. in SFC         & $\checkmark$ & 0.003            & 202.1            & 24.88 (92.2) \\
            $\hookrightarrow$ 2 Transf. in SFC         & $\checkmark$ & 16.8             & 218.9            & 25.69 (95.2) \\
            \bottomrule
        \end{tabular}
        }
    \end{table}
    \vspace{-0.25cm}

\section{Conclusions}

    We have presented SHAS, a new audio segmentation method for end-to-end ST that surpasses existing methods by a large margin in five language directions and significantly reduces the gap with manual segmentation at inference time. The proposed method is also applicable to different domains and languages. We showed that lighter versions of SHAS are still better than existing methods, hinting at potentially effective application to online settings. Future research will focus on SHAS for online ST, and on zero-shot segmentation of low-resource languages.

\section{Acknowledgments}

    This work was supported by the project Adavoice PID2019-107579RB-I00 / AEI / 10.13039/501100011033

\bibliographystyle{IEEEtran}

\bibliography{main}

\begin{thebibliography}{10}
\providecommand{\url}[1]{#1}
\csname url@samestyle\endcsname
\providecommand{\newblock}{\relax}
\providecommand{\bibinfo}[2]{#2}
\providecommand{\BIBentrySTDinterwordspacing}{\spaceskip=0pt\relax}
\providecommand{\BIBentryALTinterwordstretchfactor}{4}
\providecommand{\BIBentryALTinterwordspacing}{\spaceskip=\fontdimen2\font plus
\BIBentryALTinterwordstretchfactor\fontdimen3\font minus
  \fontdimen4\font\relax}
\providecommand{\BIBforeignlanguage}[2]{{%
\expandafter\ifx\csname l@#1\endcsname\relax
\typeout{** WARNING: IEEEtran.bst: No hyphenation pattern has been}%
\typeout{** loaded for the language `#1'. Using the pattern for}%
\typeout{** the default language instead.}%
\else
\language=\csname l@#1\endcsname
\fi
#2}}
\providecommand{\BIBdecl}{\relax}
\BIBdecl

\bibitem{attention}
A.~Vaswani, N.~Shazeer, N.~Parmar, J.~Uszkoreit, L.~Jones, A.~N. Gomez, L.~u.
  Kaiser, and I.~Polosukhin, ``Attention is {A}ll you {N}eed,'' in
  \emph{Advances in Neural Information Processing Systems}, vol.~30.\hskip 1em
  plus 0.5em minus 0.4em\relax Curran Associates, Inc., 2017.

\bibitem{mustc}
M.~A. Di~Gangi, R.~Cattoni, L.~Bentivogli, M.~Negri, and M.~Turchi,
  ``{M}u{ST}-{C}: a {M}ultilingual {S}peech {T}ranslation {C}orpus,'' in
  \emph{Proceedings of the 2019 Conference of the North {A}merican Chapter of
  the Association for Computational Linguistics: Human Language Technologies,
  Volume 1 (Long and Short Papers)}.\hskip 1em plus 0.5em minus 0.4em\relax
  Minneapolis, Minnesota: Association for Computational Linguistics, Jun. 2019,
  pp. 2012--2017.

\bibitem{mtedx}
E.~Salesky, M.~Wiesner, J.~Bremerman, R.~Cattoni, M.~Negri, M.~Turchi, D.~W.
  Oard, and M.~Post, ``{The {M}ultilingual {TED}x {C}orpus for {S}peech
  {R}ecognition and {T}ranslation},'' in \emph{Proc. Interspeech 2021}, 2021,
  pp. 3655--3659.

\bibitem{cascade_vs_end2end}
L.~Bentivogli, M.~Cettolo, M.~Gaido, A.~Karakanta, A.~Martinelli, M.~Negri, and
  M.~Turchi, ``Cascade versus {D}irect {S}peech {T}ranslation: {D}o the
  {D}ifferences {S}till {M}ake a {D}ifference?'' in \emph{Proceedings of the
  59th Annual Meeting of the Association for Computational Linguistics and the
  11th International Joint Conference on Natural Language Processing (Volume 1:
  Long Papers)}.\hskip 1em plus 0.5em minus 0.4em\relax Online: Association for
  Computational Linguistics, Aug. 2021, pp. 2873--2887.

\bibitem{matusov2007}
E.~Matusov, D.~Hillard, M.~Magimai-Doss, D.~Hakkani-Tur, M.~Ostendorf, and
  H.~Ney, ``Improving speech translation with automatic boundary prediction,''
  08 2007, pp. 2449--2452.

\bibitem{apptek2018}
E.~Matusov, P.~Wilken, P.~Bahar, J.~Schamper, P.~Golik, A.~Zeyer, J.~A.
  Silvestre-Cerda, A.~Martinez-Villaronga, H.~Pesch, and J.-T. Peter, ``Neural
  {S}peech {T}ranslation at {A}pp{T}ek,'' in \emph{Proceedings of the 15th
  International Workshop on Spoken Language Translation}, 2018, pp. 104--111.

\bibitem{iwslt2020}
E.~Ansari, A.~Axelrod, N.~Bach, O.~Bojar, R.~Cattoni, F.~Dalvi, N.~Durrani,
  M.~Federico, C.~Federmann, J.~Gu, F.~Huang, K.~Knight, X.~Ma, A.~Nagesh,
  M.~Negri, J.~Niehues, J.~Pino, E.~Salesky, X.~Shi, S.~St{\"u}ker, M.~Turchi,
  A.~Waibel, and C.~Wang, ``{FINDINGS} {OF} {THE} {IWSLT} 2020 {EVALUATION}
  {CAMPAIGN},'' in \emph{Proceedings of the 17th International Conference on
  Spoken Language Translation}.\hskip 1em plus 0.5em minus 0.4em\relax Online:
  Association for Computational Linguistics, Jul. 2020, pp. 1--34.

\bibitem{iwslt2021}
A.~Anastasopoulos, O.~Bojar, J.~Bremerman, R.~Cattoni, M.~Elbayad, M.~Federico,
  X.~Ma, S.~Nakamura, M.~Negri, J.~Niehues, J.~Pino, E.~Salesky, S.~St{\"u}ker,
  K.~Sudoh, M.~Turchi, A.~Waibel, C.~Wang, and M.~Wiesner, ``{FINDINGS} {OF}
  {THE} {IWSLT} 2021 {EVALUATION} {CAMPAIGN},'' in \emph{Proceedings of the
  18th International Conference on Spoken Language Translation (IWSLT
  2021)}.\hskip 1em plus 0.5em minus 0.4em\relax Bangkok, Thailand (online):
  Association for Computational Linguistics, Aug. 2021, pp. 1--29.

\bibitem{vad}
J.~Sohn, N.~S. Kim, and W.~Sung, ``A {S}tatistical {M}odel-{B}ased {V}oice
  {A}ctivity {D}etection,'' \emph{IEEE Signal Processing Letters}, vol.~6,
  no.~1, pp. 1--3, 1999.

\bibitem{srpol2020}
T.~Potapczyk and P.~Przybysz, ``{SRPOL}{'}s {S}ystem for the {IWSLT} 2020
  {E}nd-to-{E}nd {S}peech {T}ranslation {T}ask,'' in \emph{Proceedings of the
  17th International Conference on Spoken Language Translation}.\hskip 1em plus
  0.5em minus 0.4em\relax Online: Association for Computational Linguistics,
  Jul. 2020, pp. 89--94.

\bibitem{beyond_vad}
M.~Gaido, M.~Negri, M.~Cettolo, and M.~Turchi, ``Beyond {V}oice {A}ctivity
  {D}etection: {H}ybrid {A}udio {S}egmentation for {D}irect {S}peech
  {T}ranslation,'' in \emph{Proceedings of The Fourth International Conference
  on Natural Language and Speech Processing (ICNLSP 2021)}.\hskip 1em plus
  0.5em minus 0.4em\relax Trento, Italy: Association for Computational
  Linguistics, 12--13 Nov. 2021, pp. 55--62.

\bibitem{wav2vec2.0}
A.~Baevski, Y.~Zhou, A.~Mohamed, and M.~Auli, ``wav2vec 2.0: A {F}ramework for
  {S}elf-{S}upervised {L}earning of {S}peech {R}epresentations,'' in
  \emph{Advances in Neural Information Processing Systems}, H.~Larochelle,
  M.~Ranzato, R.~Hadsell, M.~F. Balcan, and H.~Lin, Eds., vol.~33.\hskip 1em
  plus 0.5em minus 0.4em\relax Curran Associates, Inc., 2020, pp.
  12\,449--12\,460.

\bibitem{xls-r}
A.~Babu, C.~Wang, A.~Tjandra, K.~Lakhotia, Q.~Xu, N.~Goyal, K.~Singh, P.~von
  Platen, Y.~Saraf, J.~Pino \emph{et~al.}, ``{XLS-R}: {S}elf-supervised
  {C}ross-lingual {S}peech {R}epresentation {L}earning at {S}cale,''
  \emph{arXiv preprint arXiv:2111.09296}, 2021.

\bibitem{sinclair2014}
M.~Sinclair, P.~Bell, A.~Birch, and F.~McInnes, ``{A semi-Markov model for
  speech segmentation with an utterance-break prior},'' in \emph{Proc.
  Interspeech 2014}, 2014, pp. 2351--2355.

\bibitem{upc2021}
G.~I. G{\'a}llego, I.~Tsiamas, C.~Escolano, J.~A.~R. Fonollosa, and M.~R.
  Costa-juss{\`a}, ``End-to-end {S}peech {T}ranslation with {P}re-trained
  {M}odels and {A}dapters: {UPC} at {IWSLT} 2021,'' in \emph{Proceedings of the
  18th International Conference on Spoken Language Translation (IWSLT
  2021)}.\hskip 1em plus 0.5em minus 0.4em\relax Bangkok, Thailand (online):
  Association for Computational Linguistics, Aug. 2021, pp. 110--119.

\bibitem{europarl}
J.~Iranzo-Sánchez, J.~A. Silvestre-Cerdà, J.~Jorge, N.~Roselló, A.~Giménez,
  A.~Sanchis, J.~Civera, and A.~Juan, ``Europarl-{ST}: {A} {M}ultilingual
  {C}orpus for {S}peech {T}ranslation of {P}arliamentary {D}ebates,'' in
  \emph{ICASSP 2020 - 2020 IEEE International Conference on Acoustics, Speech
  and Signal Processing (ICASSP)}, 2020, pp. 8229--8233.

\bibitem{huggingface}
T.~Wolf, L.~Debut, V.~Sanh, J.~Chaumond, C.~Delangue, A.~Moi, P.~Cistac,
  T.~Rault, R.~Louf, M.~Funtowicz, J.~Davison, S.~Shleifer, P.~von Platen,
  C.~Ma, Y.~Jernite, J.~Plu, C.~Xu, T.~L. Scao, S.~Gugger, M.~Drame, Q.~Lhoest,
  and A.~M. Rush, ``Transformers: {S}tate-of-the-{A}rt {N}atural {L}anguage
  {P}rocessing,'' in \emph{Proceedings of the 2020 Conference on Empirical
  Methods in Natural Language Processing: System Demonstrations}.\hskip 1em
  plus 0.5em minus 0.4em\relax Online: Association for Computational
  Linguistics, Oct. 2020, pp. 38--45.

\bibitem{prelayer_norm}
R.~Xiong, Y.~Yang, D.~He, K.~Zheng, S.~Zheng, C.~Xing, H.~Zhang, Y.~Lan,
  L.~Wang, and T.~Liu, ``On {L}ayer {N}ormalization in the {T}ransformer
  {A}rchitecture,'' in \emph{Proceedings of the 37th International Conference
  on Machine Learning}, ser. Proceedings of Machine Learning Research, vol.
  119.\hskip 1em plus 0.5em minus 0.4em\relax PMLR, 13--18 Jul 2020, pp.
  10\,524--10\,533.

\bibitem{gelu}
D.~Hendrycks and K.~Gimpel, ``Gaussian {E}rror {L}inear {U}nits ({GELU}s),''
  \emph{arXiv preprint arXiv:1606.08415}, 2020.

\bibitem{adam}
D.~P. Kingma and J.~Ba, ``Adam: {A} {M}ethod for {S}tochastic {O}ptimization,''
  in \emph{Proceedings of 3rd International Conference on Learning
  Representations, {ICLR}, San Diego, USA}, 2015.

\bibitem{joint-s2t}
Y.~Tang, J.~Pino, X.~Li, C.~Wang, and D.~Genzel, ``Improving {S}peech
  {T}ranslation by {U}nderstanding and {L}earning from the {A}uxiliary {T}ext
  {T}ranslation {T}ask,'' in \emph{ACL}, 2021.

\bibitem{fairseq}
C.~Wang, Y.~Tang, X.~Ma, A.~Wu, D.~Okhonko, and J.~Pino, ``fairseq {S2T}:
  {F}ast {S}peech-to-{T}ext {M}odeling with fairseq,'' in \emph{Proceedings of
  the 2020 Conference of the Asian Chapter of the Association for Computational
  Linguistics (AACL): System Demonstrations}, 2020.

\bibitem{sentencepiece}
T.~Kudo and J.~Richardson, ``{S}entence{P}iece: A simple and language
  independent subword tokenizer and detokenizer for {N}eural {T}ext
  {P}rocessing,'' in \emph{Proceedings of the 2018 Conference on Empirical
  Methods in Natural Language Processing: System Demonstrations}.\hskip 1em
  plus 0.5em minus 0.4em\relax Brussels, Belgium: Association for Computational
  Linguistics, Nov. 2018, pp. 66--71.

\bibitem{mwersegmenter}
E.~Matusov, G.~Leusch, O.~Bender, and H.~Ney, ``Evaluating {M}achine
  {T}ranslation {O}utput with {A}utomatic {S}entence {S}egmentation,'' in
  \emph{Proceedings of the Second International Workshop on Spoken Language
  Translation}, Pittsburgh, Pennsylvania, USA, Oct. 24-25 2005.

\bibitem{bleu}
K.~Papineni, S.~Roukos, T.~Ward, and W.-J. Zhu, ``{BLEU}: {A} {M}ethod for
  {A}utomatic {E}valuation of {M}achine {T}ranslation,'' in \emph{Proceedings
  of the 40th Annual Meeting on Association for Computational Linguistics},
  ser. ACL '02.\hskip 1em plus 0.5em minus 0.4em\relax USA: Association for
  Computational Linguistics, 2002, p. 311–318.

\bibitem{sacrebleu}
M.~Post, ``A {C}all for {C}larity in {R}eporting {BLEU} {S}cores,'' in
  \emph{Proceedings of the Third Conference on Machine Translation: Research
  Papers}.\hskip 1em plus 0.5em minus 0.4em\relax Belgium, Brussels:
  Association for Computational Linguistics, Oct. 2018, pp. 186--191.

\end{thebibliography}

\end{document}